\documentclass{aastex631}

\shorttitle{Integration Methods for Modeling Reactions}
\shortauthors{Johnson et al.}
\graphicspath{{./}{figures/}}
\begin{document}

\title{A Fully Explicit Integrator for Modeling Astrophysical Reactive Flows}

%\correspondingauthor{August Muench}
%\email{greg.schwarz@aas.org, gus.muench@aas.org}

\author[0009-0005-5534-7495]{Parker Johnson}
\affiliation{Dept. of Physics and Astrophysics, University of North Dakota, Grand Forks, ND, 58202, USA}
\author[0000-0001-8401-030X]{Michael Zingale}
\affiliation{Dept. of Physics and Astronomy, Stony Brook University, Stony Brook, NY 11794-3800, USA}
\author[0000-0003-3603-6868]{Eric T. Johnson}
\affiliation{Dept. of Physics and Astronomy, Stony Brook University, Stony Brook, NY 11794-3800, USA}
\author[0000-0001-5961-1680]{Alexander Smith}
\affiliation{Dept. of Physics and Astronomy, Stony Brook University, Stony Brook, NY 11794-3800, USA}
\author[0000-0003-4425-7097]{Kyle E.~Niemeyer}
\affiliation{School of Mechanical, Industrial, and Manufacturing Engineering, Oregon State University, Corvallis, OR 97331, USA}

\begin{abstract}

Simulating complex astrophysical reacting flows is computationally expensive---reactions are stiff and typically require implicit integration methods.  The reaction update is often the most expensive part of a simulation, which motivates the exploration of more economical methods. In this research note, we investigate how the explicit Runge--Kutta--Chebyshev (RKC) method performs compared to an implicit method when applied to astrophysical reactive flows. These integrators are applied to simulations of X-ray bursts arising from unstable thermonuclear burning of accreted fuel on the surface of neutron stars.  We show that the RKC method performs with similar accuracy to our traditional implicit integrator, but is more computationally efficient when run on CPUs.
\end{abstract}


\section{Introduction} \label{sec:intro}

In an X-ray burst (XRB), a neutron star accretes He or H/He from a companion star building up a thin layer on its surface.  The strong gravitational acceleration compresses this material to the point of thermonuclear runaway, producing a brief flash of X-rays \citep{galloway:2017}. Observations suggest that the burning begins locally and spreads across the neutron star as a flame \citep{bhattacharyya:2007}.

Modeling XRB flames can be challenging due to the difference in burning and hydrodynamic timescales. Traditionally, explicit integration methods are applied to hydrodynamics while implicit integration methods are applied to the stiff reaction systems \citep{hujeirat:2001}. For an
ordinary differential equation (ODE) system,
\begin{equation}
   \dot{\mathbf{y}} = \mathbf{f}(t, \mathbf{y}(t)) \;,
\end{equation}
a first-order implicit discretization over a timestep $\Delta t$ is:
\begin{equation}
    \mathbf{y}^{n+1} = \mathbf{y}^n + \Delta t \mathbf{f}(t^{n+1}, \mathbf{y}^{n+1}) \;.
\end{equation}
For small $\Delta t$, the change in the solution, $\Delta \mathbf{y}$, will be small as well, allowing the system to be Taylor-expanded around $(t^{n+1}, \mathbf{y}^{n+1})$. This requires the Jacobian matrix, $\textbf{J} = \partial 
\textbf{f} / \partial \textbf{y}$ to be computed and stored
every timestep leading to computationally expensive linear system solves.  In contrast, explicit methods do not need the Jacobian, instead constructing the update from the current solution.  Explicit methods can use less memory,  but they can be unstable when applied to the reacting system. In this paper, we study explicit methods with increased stability applied to the reacting system, with the goal of conserving memory and increasing the overall performance for large-scale simulations.

\citet{niemeyersung} showed that the Runge--Kutta--Chebyshev (RKC) method \citep{rkc} can be an efficient integrator when applied to chemical kinetics in terrestrial reactive-flow simulations. RKC is based on explicit Runge--Kutta methods, using the Chebyshev formulas to add stages
to the integration to increase its stability region.
%(stability increases quadratically with the number of stages).
At each step, an estimate of the spectral radius is needed, which is computed internally in RKC using the power method without constructing
the Jacobian.

% RKC is appropriate for solutions of modest accuracy in mildly stiff problems with eigenvalues
% of Jacobians that are close to the negative real axis. For these problems, RKC has the 
% advantages of explicit one-step methods and extremely low storage.

We implement RKC for astrophysical nuclear reactions in the Castro hydrodynamic code \citep{castro,castro_joss}. Castro uses the AMReX adaptive mesh library \citep{amrex_joss} and a general equation of state and nuclear reaction network, diffusion, and gravity.
The advective update is fully explicit, using an unsplit piecewise parabolic method \citep{millercolella:2002}.
Several methods for coupling hydrodynamics and reactions are available, here we consider Strang-splitting \citep{strang:1968} and the simplified spectral deferred corrections (SDC) approach \citep{castro_simple_sdc}.
In all cases, the reaction system updates species and energy according to the reaction
rates.

In the next section, we compare RKC and the implicit VODE integrator \citep{vode} for
simulations of XRBs.  We look at both CPU and GPU performance and consider both the Strang-split coupling and the simplified-SDC method.

\section{Results} \label{sec:simulations}

We use the Castro {\tt flame\_wave} setup \citep{eiden:2020,harpole:2021} for our XRB flame model.
The domain is $0.256 \,\mathrm{km}\times 1.024\,\mathrm{km}$, with a base grid of $512\times 128$ and two levels of refinement (jumps of $4\times$, then $2\times$), giving a 25~cm resolution overall.
Reactions are modeled using the 13 isotope {\tt aprox13} network \citep{aprox13}.
We ran eight simulations, each on eight nodes of the NERSC
Perlmutter machine.
CPU runs used 128 MPI tasks each with eight OpenMP threads;  GPU runs used 32 NVIDIA A100 GPUs.
For the RKC integrator
%, the spectral radius was computed via the power method and 
we found it critical to scale the energy to be $\mathcal{O}(1)$.

\begin{figure}[t]
\centering
\plotone{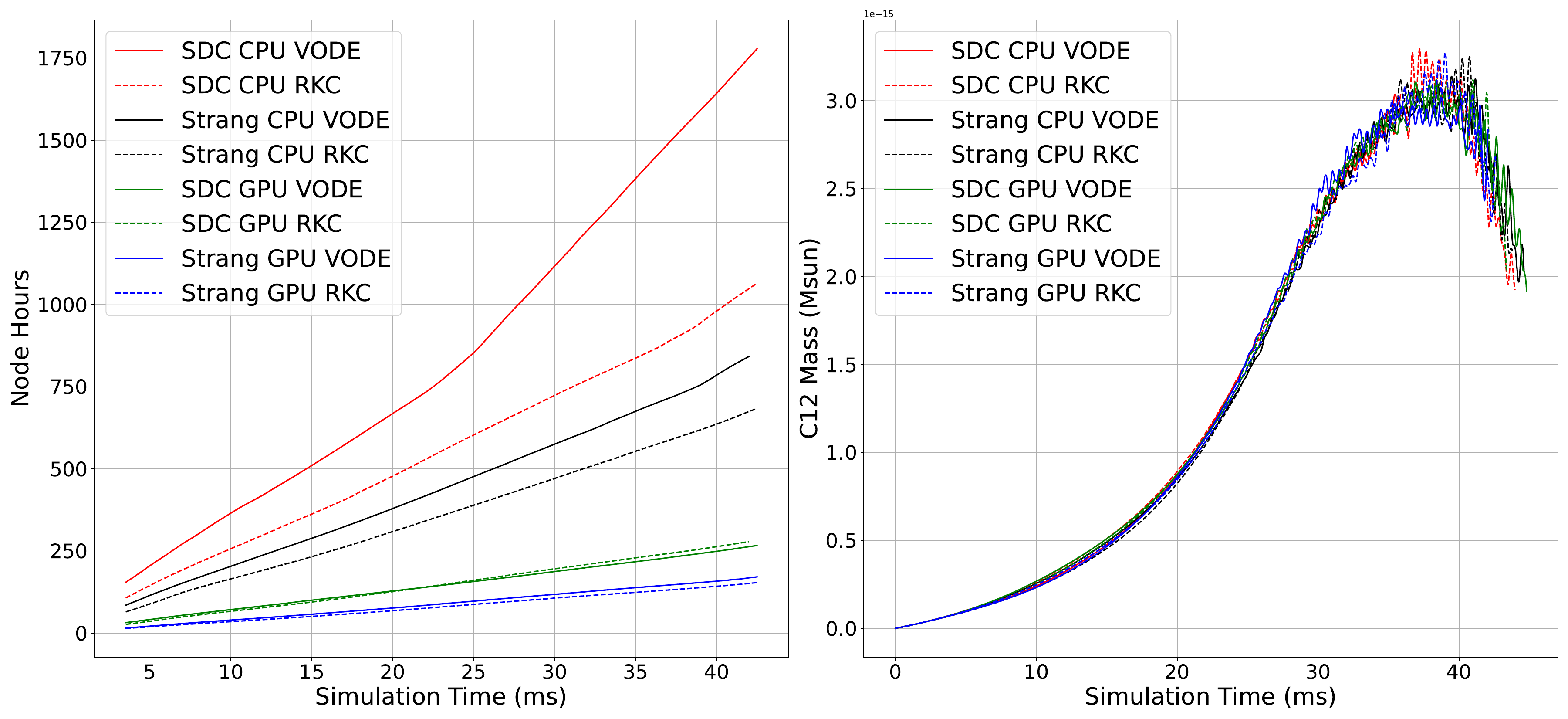}
\caption{\label{fig:results} Number of node hours and carbon mass as a function of time for the different integrators and time-integration method on CPU/GPU.}
\end{figure}

Figure~\ref{fig:results} shows the computational cost as a function of simulation time for each simulation.
We see that RKC outperforms VODE when using Strang and simplified-SDC integration on CPUs. 
On GPUs, the performance between the RKC and VODE is almost the same, regardless of the integration method. 
This differs from the findings of \citet{niemeyersung,stonedavis} and may be due to the differing costs in evaluating the nuclear reaction right-hand side function or simply because our VODE implementation has been optimized more than RKC.
As expected, the GPU runs overall are much faster than the corresponding CPU runs. 
%The amount of node hours decrease by 32.2 percent when using RKC with SDC on CPUs, decrease by 18.2 percent when using RKC with Strang on CPUs, decrease by 10.1 percent when using RKC with Strang on GPUs, but increases by 3.6 percent when using RKC with SDC on GPUs.
The right panel of the figure shows the total ${}^{12}\mathrm{C}$  mass from burning ${}^{4}\mathrm{He}$.  
All simulations show the same trend: an initial build up of carbon and then a drop as it is consumed to produce heavier elements.
We also see that for this problem, the simplified-SDC method does not improve the efficiency of this simulation, consistent with  \citep{zhi2023}.

%%\section{Summary}

Our results show that a fully explicit integration method for reactions can be more efficient than an implicit method in modeling XRBs.
%There was a significant decrease in the amount of node hours needed to run the simulations when using CPUs. %The implication of these findings is the ability to save hundreds of node hours by simply implementing the explicit RKC integrator.  
Finally, we note that a similar study was done looking at He detonations, but there RKC struggles to perform as well as VODE, consistent with findings for more-stiff terrestrial combustion simulations~\citep{niemeyersung}.
This is likely because the reactions in the detonation are much more stiff and require many more stages, so the cost of evaluating the right-hand side of the reaction ODE system dominates. 
Future work will explore larger networks as well as higher-order stabilized explicit integrators, like the ROCK family \citep{rock4}.

\begin{acknowledgments}

Castro is available at \href{https://github.com/AMReX-Astro/Castro}{http://github.com/AMReX-Astro/Castro}.  This research used the DOE National Energy Research Scientific Computing Center and was supported by DOE
Office of Nuclear Physics grant DE-FG02-87ER40317
and NSF grant PHY-2243856. 

\end{acknowledgments}


\vspace{5mm}
\facilities{NERSC}


\software{AMReX \citep{amrex_joss};
          Castro \citep{castro, castro_joss},
          matplotlib \citep{Hunter:2007},
          VODE \citep{vode}}

\bibliography{refs}{} 
\bibliographystyle{aasjournal}



\end{document}